\documentclass[aps,prc,nofootinbib]{revtex4}
\usepackage{eurosym}
\usepackage{amsmath,amssymb}
\usepackage{graphicx}
\usepackage{dcolumn}
\usepackage{bm}
\usepackage{color}
\usepackage{multirow}
\usepackage{float}
\usepackage{array}
\usepackage{dcolumn}
\usepackage{booktabs}
\usepackage[detect-all]{siunitx}
\usepackage{makecell}
\usepackage{xcolor}
\usepackage{yfonts}

\graphicspath{ {/Users/Nastia/Desktop/research_dis/papers_writing/xenes/nov_5} }

\newcolumntype{P}[1]{>{\centering\arraybackslash}p{#1}}
\begin{document}

\title{Bound states of $^{9}_{\phi}$Be and $^{6}_{\phi\phi}$He nuclei with $\phi$+$\alpha$+$\alpha$ and $\phi$+$\phi$+$\alpha$ cluster models
}
\author{I. Filikhin$^1$, R. Ya. Kezerashvili$^{2,3,4}$, and B. Vlahovic$^1$}
\affiliation{\mbox{$^{1}$North Carolina Central University, Durham, NC, USA} \\
$^{2}$New York City College of Technology, The City University of New York,
Brooklyn, NY, USA\\
$^{3}$The Graduate School and University Center, The City University of New
York, New York, NY, USA\\
$^{4}$Long Island University, Brooklyn, NY, USA}

\begin{abstract}

We investigate the $^{9}_{\phi}$Be and $^{6}_{\phi\phi}$He $\phi$ mesic nuclei within the framework of the three-body cluster model 
as the $\phi$+$\alpha$+$\alpha$ and $\phi$+$\phi$+$\alpha$ systems, using the Faddeev formalism in configuration space. The $\phi$-$\alpha$ potential is determined through a folding procedure of the HAL QCD $\phi$-$N$ interaction in the $^4S_{3/2}$ channel with the matter distribution of $^4$He. The phenomenological $\alpha$-$\alpha$ and $\phi$-$\phi$ potentials are taken from the literature. Additionally, we construct a Wood-Saxon (WS) type interaction to simulate the $\phi$-$\alpha$ potential, also taken from the literature, based on an effective Lagrangian approach that includes $K\bar{K}$ meson loops in the $\phi$-meson self-energy. A comparison of binding energies obtained for both types of the $\phi$-$\alpha$ interactions reveals qualitative agreement. 
We predict the binding energy for the $^{9}_{\phi}$Be and $^{6}_{\phi\phi}$He $\phi$ mesic nuclei as the mirror $\phi$+$\alpha$+$\alpha$ and $\phi$+$\phi$+$\alpha$ systems in the range of 1-11 MeV and 3-10 MeV, respectively. The range of values of the binding energies relies on the choice of the WS $\phi$-$\alpha$ interaction parameters.

\end{abstract}

\maketitle
\date{\today }

\section{Introduction}
\label{intro}

Meson-nucleus systems bound by attractive strong interactions
are very interesting objects and have received considerable experimental and theoretical interest in the last few decades \cite{Hayano2010,Paryev2017,Krein2018,RKezNNNK}. Vector mesons interaction with nuclei 
have been studied for quite a long time, especially in conjunction with the expected restoration of chiral symmetry in nuclear matter.

The $\phi$ meson is a vector meson composed of a strange quark and anti-strange quark. $\phi$-mesic nuclei that are strongly interacting
exotic many-body systems have 
recently received renewed interest. The key issues here are firstly whether $\phi$-meson is indeed bound to nuclei, secondly by how much is the bounding energy of states and what are the properties of such states, thirdly on the potential modifications of the $\phi$ meson in nuclear matter and their experimental detection \cite{Gubler2021}.
Theoretically, the properties of the $\phi$ meson in nuclear matter have been discussed based on hadronic models  \cite{Weise1998,Oset2001,Cabrera2003,Cabrera2017}, the QCD sum rules \cite{QCDF1992,QCDF1997,QCDF2015,QCDF2016,QCDF2022}, and the effective Lagrangian approach \cite{CoPL,Co17}.

For a description of $p$-shell hypernuclei various theoretical approaches, e.g., the shell model \cite{Shell1,Shell2,Shell3,MShell1,MShell2,MShell3,MShell4}, \textit{ab initio} no-core shell model \cite{NoShell1,MShell5,Wirth,MShell6,MShell7}, a mean field model based on realistic 2-body baryon interactions \cite {Hiyama16}, and cluster models \cite{Motoba832,Motoba83,Motoba85,HiyamaCluster2001,HiyamaPPNP2009} were developed. It is well known that $\alpha$ clustering plays a crucial role in light nuclei. In addition to the light nuclei, the existence
of $\alpha$ clustering in the medium-mass nuclei and $\alpha$ matter was studied, see, for example, \cite{Okada2024,Clark2024}. The study \cite{Okada2024} demonstrated that although the $j j$-coupling shell model wave function dominates around the surface region of $^{48}$Ti, the $\alpha$ clustering is important in the tail region of the wave function.
There is a considerable number of  studies of $p$-shell nuclei such as $%
_{\Lambda \Lambda }^{6}$He, $_{\Xi \Xi }^{6}$He, $_{\Omega \Omega }^{6}$He,
$_{\Lambda }^{9}$Be, $_{\Xi }^{9}$Be within three-body cluster models \cite%
{Motoba832,Motoba83,Motoba85,HiyamaCluster2001,HiyamaPPNP2009,Fujiwara2004,Suslov2004,IgorF2004,Hiyama2012FBS,Wu2020,HiyamaRev2018}. In this paper, we suggest studying $\phi$-mesic nuclei $_{\phi \phi }^{6}$He and $_{\phi}^{9}$Be in the framework of the three-body cluster model using the Faddeev equations in configuration space. Such an approach requires $\phi$-$\alpha$ potential.

The recent ALICE Collaboration measurement of the $\phi N$ correlation function \cite{ALICE2021} led
to a determination of the $\phi N$ channel scattering length with a large real part corresponding to an attractive interaction.
This is the first experimental evidence of the strong attractive interaction between a proton
and a $\phi$ meson.

The interaction between a $\phi$ meson and nucleon plays a significant role in understanding various phenomena in nuclear and particle physics. This potential describes the force between the $\phi$ meson and 
nucleons within the nucleus. In 2022 the first results on the interaction between the $\phi$-meson and the nucleon are presented based on the $(2+1)$-flavor lattice QCD simulations with nearly physical quark masses \cite{Lyu22}. Using the HAL QCD method, based on
the spacetime correlation of the $\phi$-$N$ system in the spin 3/2 channel authors suggested fits of the lattice QCD potential
by using two different functional forms: one fit is motivated by the two-pion exchange tail and the other one is a purely phenomenological Gaussian form. All fits provide an equally good result \cite{Lyu22}. Also authors \cite{Lyu22} found that the simple fitting functions such as the
Yukawa form \cite{G2001} cannot reproduce the lattice data.


This article has two foci: i. Construction of a $\phi$-$\alpha$ interaction based on the HAL QCD $\phi$-$N$ potential; ii. study the possible formation of $^{9}_{\phi}$Be and $^{6}_{\phi\phi}$He $\phi$ mesic nuclei as the $\phi+\alpha+\alpha$ and $\phi+\phi+\alpha$ system, respectively, in the framework of a three-particle cluster model. This study is carried out in the framework of Faddeev equations in configuration space.

The paper is organized as follows. In Sec. \ref{Fadd}, we present the Faddeev equations formalism in
configuration space for the description of a three-particle system when two particles are identical. The $\alpha$-$\alpha$, $\phi$-$N$, $\phi$-$\phi$, and $\phi$-$\alpha$  interaction potentials are discussed in Sec. \ref{Interactions}. In Sec \ref{Calculations}, we propose the $\phi$-$\alpha$ potential obtained based on the Wood-Saxson fit of the folding of the HAL QCD $\phi$-$N$ interaction \cite{Lyu22} and present results of
numerical calculations for
$^{9}_{\phi}$Be and $^{6}_{\phi\phi}$He $\phi$ mesic nuclei. The concluding remarks follow in Sec. \ref{sec:6}.

\section{Faddeev equations for three-body cluster systems}

\label{Fadd}
The 
$^{9}_{\phi}$Be and $^{6}_{\phi\phi}$He $\phi$ mesic nuclei as the
$\phi$+$\alpha$+$\alpha$ and $\phi$+$\phi$+$\alpha$ 
in a cluster model, represent three-particle systems.
The three-body problem can be solved in the framework of the Schr\"{o}dinger
equation or using the Faddeev approach in the momentum \cite{Fad,Fad1} or
configuration \cite{Noyes1968,Noyes1969,Gignoux1974,FM,K86} spaces. The Faddeev
equations in the configuration space have different form depending on the
type of particles and can be written for three nonidentical particles, 
three particles when two are identical, and 
three identical particles. 
The identical
particles have the same masses and quantum numbers. We consider the $\phi$+$\phi$+$\alpha$ and $\phi$+$\alpha$+$\alpha$ systems in the framework of the three-body cluster model to seek possible bound states of the $^6_{\phi\phi}$He and $^9_\phi$Be $\phi$ mesic nuclei. 
In the Faddeev
method in configuration space, alternatively, to the finding the wave
function of the three-body system using the Schr\"{o}dinger equation, the
total wave function is decomposed into three components \cite%
{Noyes1968,FM,K86}:
\begin{equation}
\Psi (\mathbf{x}_{1},\mathbf{y}_{1})=\Phi_{1}(\mathbf{x}_{1},\mathbf{y}%
_{1})+\Phi_{2}(\mathbf{x}_{2},\mathbf{y}_{2})+\Phi_{3}(\mathbf{x}_{3},%
\mathbf{y}_{3}).  \label{P}
\end{equation}%
Each Faddeev component corresponds to a separation of particles into
configurations $(kl)+i$, $i\neq k\neq l=1,2,3$. The Faddeev components are
related to its own set of the Jacobi coordinates ($\mathbf{x}_{i}$, $\mathbf{%
y}_{i}$), $i=1,2,3$. There are three sets of Jacobi coordinates. The total
wave function can be presented by the coordinates of one of the sets as is
shown in Eq. (\ref{P}) for the set $i=1$. The mass-scaled Jacobi coordinates
$\mathbf{x}_{i}$ and $\mathbf{y}_{i}$ are expressed via the particle
coordinates $\mathbf{r}_{i}$ and masses $m_{i}$ in the following form:
\begin{equation}
\mathbf{x}_{i}=\sqrt{\frac{2m_{k}m_{l}}{m_{k}+m_{l}}}(\mathbf{r}_{k}-\mathbf{%
r}_{l}),\qquad \mathbf{y}_{i}=\sqrt{\frac{2m_{i}(m_{k}+m_{l})}{%
m_{i}+m_{k}+m_{l}}}(\mathbf{r}_{i}-\frac{m_{k}\mathbf{r}_{k}+m_{l}\mathbf{r}%
_{l})}{m_{k}+m_{l}}).  \label{Jc}
\end{equation}%
In Eq. (\ref{P}), the components depend on the corresponding coordinate set
which are expressed in terms of the chosen set of mass-scaled Jacobi
coordinates. The orthogonal transformation between three different sets of
the Jacobi coordinates has the form:
\begin{equation}
\left(
\begin{array}{c}
\label{tran}\mathbf{x}_{i} \\
\mathbf{y}_{i}%
\end{array}%
\right) =\left(
\begin{array}{cc}
C_{ik} & S_{ik} \\
-S_{ik} & C_{ik}%
\end{array}%
\right) \left(
\begin{array}{c}
\mathbf{x}_{k} \\
\mathbf{y}_{k}%
\end{array}%
\right) ,\ \ C_{ik}^{2}+S_{ik}^{2}=1, \quad k\neq i,
\end{equation}%
where
\begin{equation*}
C_{ik}=-\sqrt{\frac{m_{i}m_{k}}{(M-m_{i})(M-m_{k})}},\quad S_{ik}=(-1)^{k-i}%
\mathrm{sign}(k-i)\sqrt{1-C_{ik}^{2}}.
\end{equation*}%
Here, $M$ is the total mass of the system. Let us definite the
transformation $h_{ik}(\mathbf{x},\mathbf{y})$ based on Eq. (\ref{tran}) as
\begin{equation}
h_{ik}(\mathbf{x},\mathbf{y})=\left(C_{ik} \mathbf{x}+ S_{ik}\mathbf{y},
-S_{ik}\mathbf{x}+ C_{ik}\mathbf{y} \right).  \label{Trans}
\end{equation}
The transformation (\ref{Trans}) allows to write the Faddeev equations in
compact form. The components $\Phi_i(\mathbf{x}_{i},\mathbf{y}_{i})$ satisfy
the Faddeev equations \cite{FM} and can be written in the coordinate
representation as:
\begin{equation}
(H_{0}+V_{i}(|C_{ik}\mathbf{x}|)-E)\Phi_i(\mathbf{x},\mathbf{y})=-V_{i}(|C_{ik}%
\mathbf{x}|)\sum_{l\neq i}\Phi_l(h_{il}(\mathbf{x},\mathbf{y})).  \label{e:1}
\end{equation}
Here, 
$H_{0}=-(\Delta _{\mathbf{x}}+\Delta _{\mathbf{y}})$ is the
kinetic energy operator with $\hbar ^{2}=1$ and $V_{i}(|\mathbf{x}|)$ is the
interaction potential between the pair of particles $(kl)$, where $k,l\neq i$.

\begin{figure}[ht]
\begin{center}
\includegraphics[width=15pc]{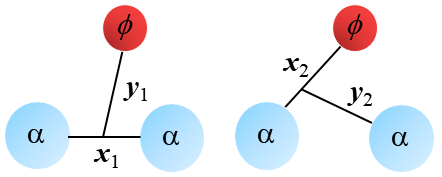}

\end{center}
\caption{Schematics for Jacobi coordinates for the $\phi$+$\alpha$+$\alpha$ system ($^{9}_{\phi}$Be
nuclei). The coordinates correspond to two rearrangements: ($\alpha\alpha$)$\phi$ and ($\phi\alpha$)$\alpha$. The two $\alpha$ particles are to be symmetrized. For the $\phi$+$\phi$+$\alpha$ ($^{6}_{\phi\phi}$He nuclei) system one could replace the $\alpha$-particles by two $\phi$-mesons and the $\phi$-meson with the $\alpha$-particle.
}
\label{fig:03}
\end{figure}

The system of equations, Eqs. (\ref{e:1}), written for three nonidentical particles can be reduced to a simpler form
for a case of two identical particles. Within the cluster model, the $^{9}_{\phi}$Be and $^{6}_{\phi\phi}$He nuclei can be treated as a three-particle system with two identical $\alpha$ particles or two $\phi$ mesons, respectively. The formulation of the Faddeev equations for three particles within the 
model with two identical
$\phi$ mesons or $\alpha$ particles can be considered as a starting point
for the study of the $^6_{\phi\phi}$He and $^9_\phi$Be $\phi$ mesic nuclei. Even if one considers the Coulomb interaction in the $\phi$+$\alpha$+$\alpha$ system it still should be described within the two identical particle model. The system of Eqs. (\ref{e:1}) can be
reduced to a simpler form for a case of two identical particles.
The Faddeev
equations in configuration space for a three-particle system with two identical
particles are given in our previous studies \cite{Kez2017,Kez2018PL,KezPRD2020}%
. Figure \ref{fig:03} depicts schematics for Jacobi coordinates for three particles when two particles are identical. In this case, for the bosonic particles, the total wave function of the
system is decomposed into the sum of the Faddeev components $\Phi_1$ and $%
\Phi_2$ corresponding to the $(\alpha\alpha)\phi$ and $(\phi\alpha)\alpha$ or $(\phi\phi)\alpha$ and $(\phi\alpha)\phi$ types of rearrangements for the $\phi$+$\alpha$+$\alpha$ and $\phi$+$\phi$+$\alpha$, respectively:
\begin{equation*}
\Psi =\Phi_1+\Phi_2+P\Phi_2,
\end{equation*}%
where $P$ is the permutation operator for two identical bosons.
Therefore, the set of the Faddeev equations (\ref{e:1}) is rewritten as
follows \cite{K86}:
\begin{equation}
\begin{array}{l}
{(H_{0}+V_{\alpha\alpha}-E)\Phi_1=-V_{\alpha\alpha}(\Phi_2+P\Phi_2)}, \\
{(H_{0}+V_{\phi\alpha}-E)\Phi_2=-V_{\phi\alpha}(\Phi_1+P\Phi_2)}.%
\end{array}
\label{GrindEQ__1_}
\end{equation}
In Eqs. (\ref{GrindEQ__1_}) that are written for the $\phi$+$\alpha$+$\alpha$ system, ${V_{\alpha\alpha}}$ and ${V_{\phi\alpha}}$ are the interaction
potentials between identical and nonidentical particles, respectively, when the Coulomb interaction between $\alpha$ particles is ignored. To acquire the overall interactions in the $\phi$+$\alpha$+$\alpha$ system, we must consider
the inclusion of the Coulomb potential as well. In Ref. \cite{KezJPG2024} are given Faddeev equations with two identical particles considering the Coulomb potential.
Below we consider a $s$-wave model for the systems. The details of our method for the solution of the system of differential equations (\ref{GrindEQ__1_}) are given in \cite{KezPRD2020,KezJPG2024,KezJPG2016}.

\section{Interaction Potentials}

\label{Interactions}

Investigations into the properties of the $\phi$+$\alpha$+$\alpha$ and $\phi$+$\phi$+$\alpha$ systems within a nonrelativistic potential model requires $\phi$-$\alpha$, $\alpha$-$\alpha$, and $\phi$-$\phi$ interaction potentials.

\textbf{$\alpha$-$\alpha$ interaction.} The interaction between two $\alpha$ particles is expressed as a combination of nuclear and Coulomb components:

\begin{equation}
\label{AL}
V_{\alpha\alpha}(r) = V_n(r) + V_C(r).
\end{equation}
Generally there are two approaches for description of the nuclear part of (\ref{AL}). The $\alpha$-$\alpha$ potential, which reproduces the observed
$\alpha$-$\alpha$ scattering phase shift and the ground state of $^{8}$Be, and with
the $\alpha$-$\alpha$ orthogonality condition model \cite{OCM}.
A cluster model when the Pauli principle between nucleons belonging to two $\alpha$ clusters is taken into account by the orthogonality condition \cite{OCM} and including the Pauli exclusion operator into the Hamiltonian of systems was employed widely, see reviews \cite{Hiyama2012FBS,HiyamaRev2018} and references herein. 
On the other hand, a nuclear part of (\ref{AL}) is typically described using various phenomenological local potential models, such as the double Gaussian Ali-Bodmer potential \cite{AliBodmer}, Morse potential \cite{Morse1929}, double Hulthen potential \cite{Hulthen2016}, or the Malfliet-Tjon potential \cite{Malfliet1969}. Suggested over 60 years ago potential \cite{AliBodmer} 
has been widely used for calculations of nuclei binding energies in the framework of the cluster model, see, for example, Refs. \cite{Jibuti1978,KezNP1984,KezYad1992,Igor2000,fed1,suslov2004,suslov2005,fed2,fed3,Ishikawa1,Ishikawa2,Ishikawa3}. With its four
parameters chosen to fit scattering data in the leading states $L = 0$, 2, 4 of angular momentum up to 24 MeV, this interaction
consists of an $L$-dependent inner repulsive Gaussian term and
an $L$-independent outer attractive Gaussian term.
In our calculations, we adopt a four-parameter double Gaussian potential \cite{AliBodmer}:
\begin{equation}
V_n(r) = V_r e^{-\mu_r^2 r^2} - V_a e^{-\mu_a^2 r^2},
\label{alfaalfa}
\end{equation}
where $V_r$ and $V_a$ represent the strengths of the repulsive and attractive parts of the potential in MeV, 
respectively. $\mu_r$ and $\mu_a$ denote their corresponding inverse ranges in fm$^{-1}$. 
Note that the remarkable similitude of the many-body problems of $\alpha$ matter and liquid
$^4$He is demonstrated in Ref. \cite{Clark2024} by showing a comparison between $\alpha$-$\alpha$  interaction \cite{AliBodmer} with $L = 0$ and the Aziz atom-atom interaction \cite{Aziz1979} in liquid $^4$He.


\textbf{$\phi$-$N$ interaction.} In Ref. \cite{Lyu22} the interaction between the $\phi$ meson and the nucleon is studied based on the
($2+1$)-flavor lattice QCD simulations with nearly physical quark masses. Authors found that the $\phi N$ correlation function
is mostly dominated by the elastic scattering states in the
$^{4}S_{3/2}$ channel without significant effects from the two-body $\Lambda K(^{2}D_{3/2})$ and $\Sigma K(^{2}D_{3/2})$ and the three-body open channels
including $\phi N \rightarrow {\Sigma^{*}K,\Lambda(1405)K} \rightarrow {\Lambda\pi K, \Sigma\pi K}$. The fit of the lattice QCD potential by the sum of two Gaussian and the two-pion exchange tail at long distance with an overall strength proportional to $m_{\pi}^{4n}$ \cite{Kreinm4} has the following functional form \cite{Lyu22}:
\begin{equation}
V_{\phi N}(r)=a_{1}e^{-r^{2}/b_{1}^{2}}+a_{2}e^{-r^{2}/b_{2}^{2}}+a_{3}m_{\pi
}^{4}F(r,b_{3})\left( \frac{e^{-m_{\pi }r}}{r}\right) ^{2},
\label{HALQCD}
\end{equation}
with the Argonne-type form factor \cite{Wiringa95}
\begin{equation}
F(r,b_{3})=(1-e^{-r^{2}/b_{3}^{2}})^{2}.
\label{Ffactor}
\end{equation}
For comparison the lattice QCD $\phi N$ potential is also parameterized using three Gaussian \cite{Lyu22}:
\begin{equation}
V_{G\phi N}(r)=\sum_{j=1}^{3}a_{j}\exp \left[ -\left( \frac{r}{b_{j}}\right)^{2}\right].
\label{phiN}
\end{equation}
Although both fit provide an equally good result in reproducing the lattice data \cite{Lyu22}, below we perform calculations with both the  lattice QCD $V_{\phi N}$ potential with a two-pion exchange tail  and a purely phenomenological sum of three Gaussian,  $V_{G\phi N}$, potential. Parameters for these potentials are given in Table \ref{tpp}. Let us mention although the HAL QCD $\phi$-$N$ potential in $^{4}S_{3/2}$ channel with the maximal spin 3/2 is found to be attractive for all distances and reproduces a two-pion exchange tail at long distances \cite{ALICE2021,Lyu22}, no bound state is found with this interaction for $\phi N$ and $\phi NN$ systems \cite{FKVPRD2024}. Thus, the HAL QCD $\phi N$ potential in the
 $^{4}S_{3/2}$ channel  does not provide enough attractiveness to support either the $\phi N$ or $\phi NN$ bound states. 

\textbf{$\phi$-$\phi$ interaction.} We use the phenomenological
$\phi$-$\phi$ potential from Ref. \cite{Bel2008,Sofi} in the form of a sum of two Yukawa terms:
\begin{equation}
\label{pp} V_{\phi \phi}(r)=V_{1}\frac{e^{-\mu_{1} r} }{r}-V_{2}\frac{e^{-\mu_{2} r} } {r}.
\end{equation}
The parameters of this potential were fixed by the position and
width of the $f_{2}$(2010) resonance which has only one decay channel
into two $\phi$-mesons \cite{Bel2008,Sofi}.

\textbf{$\phi$-$\alpha$ interaction.} Meson-nucleus systems bound by attractive interactions
are strongly interacting exotic many-body systems. The study of the  $^6_{\phi\phi}$He and $^9_\phi$Be $\phi$ mesic nuclei within a three-body cluster model needs a $\phi$-$\alpha$ interaction potential. This potential might be approximated by “folding” the $\alpha$-particle nuclear density distribution with the $\phi$-$N$ interaction.

In the single folding model the $\phi$-$\alpha$ potential, $V_{\phi\alpha}^{F}(r)$, can be obtained as \cite{Satchler1983}:
\begin{equation}
\label{folding}
V_{\phi \alpha }^{F}(r)=\int \rho (\mathbf{x)}V_{\phi N}(\mathbf{r-x)}d\mathbf{x},
\end{equation}
where  $\rho (\mathbf{x)}$ is the density of nucleons in $^{4}$He and $\left\vert \mathbf{r-x}\right\vert $ is the distance between the $\phi$ meson and nucleon. Considering the central symmetry of the $V_ {\phi N}(r)$ potential and density $\rho(r)$,  expression (\ref{folding}) reads
\begin{equation}
\label{2Dintegral}
V_{\phi\alpha}^{F}(r)=4\int_{-1}^12\pi du\int_0^{\infty}dx \rho(x)V_{\phi N}(\sqrt{x^2+r^2-2xru})x^2.
\end{equation}

At large distances between $\alpha$ and $\phi$, the clustering is described as $\phi +(NNNN)$.
In the region near the alpha cluster and inside, the clustering must include different combinations $(\phi NNN)+N$,
$(\phi NN )+(NN)$, $(\phi N)+(NNN)$.  Taking into account that the $\phi $ meson does not make bound states in the subsystems,
we assume that the $\phi+(NNNN)$ clusterization is dominating. Thus, we assume that  the folding potential is an appropriate approach for the
$\phi$-$\alpha$ interaction including the near and in regions. Let us mention  that the three Gaussian approximation of the HAL QCD potential is very instrumental, which allows evaluation (\ref{2Dintegral}) for the Gaussian matter distribution in the analytical form, as shown in Appendix A.

\begin{figure}[ht!]
\begin{center}
\includegraphics[width=16pc]{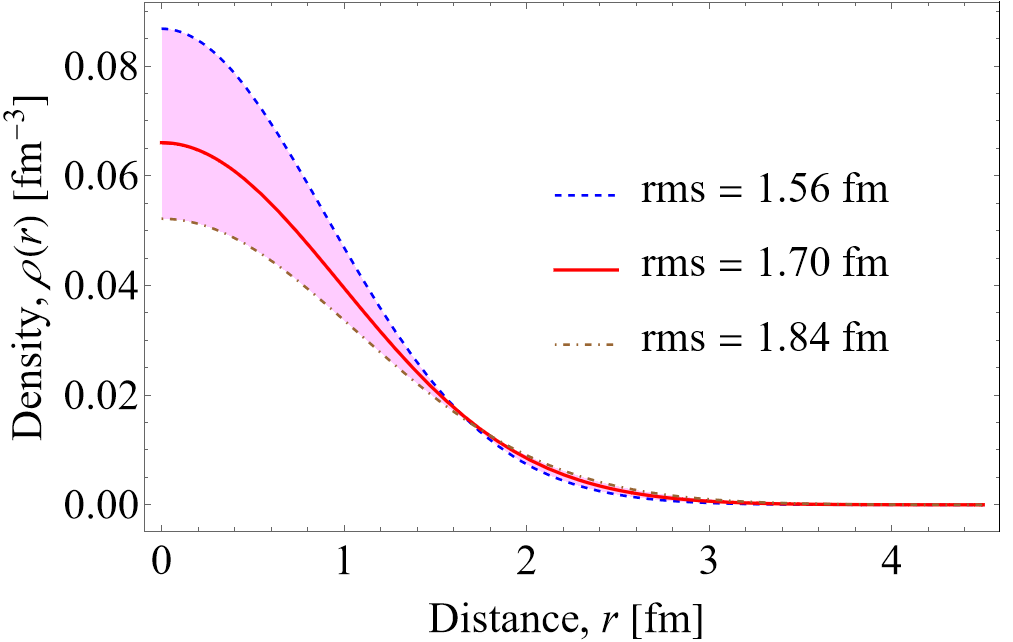}
\end{center}
\caption{The matter density distribution $\rho(r)$ in $^4$He that corresponds to the different $rms$ radii. The solid curve corresponds to the experimental $rms = 1.70$ fm \cite{Wang2024}. The shaded area corresponds to the experimental uncertainty of $\pm 0.14$ fm for $rms$ radii given by statistics \cite{Wang2024}.
}
\label{fig:05}
\end{figure}

The root-mean-square ($rms$) radius is an important and basic property for any composite subatomic system. For $^4$He both the matter and charge $rms$ radii were measured. A matter radius is related to both the proton and the neutron distributions inside a nucleus, whereas nuclear charge radius primarily connected to the proton distribution. The average $rms$ charge radius of $^4$He from electron elastic scattering experiments is 1.681(4) fm \cite{Sick2008}. While 
combined analysis gave the average $rms$ charge radius of  $^4$He  to be 1.6755(28) fm \cite{Marinova}. The result of precise measurements of  $^4$He $rms$ charge radius with the technique of muon-atom spectroscopy gives 1.67824(13) fm \cite{Krauth}. Recently, by examining the near-threshold $\phi$-meson photoproduction data of the LEPS Collaboration \cite{HiraiwaLEPS}, the $rms$ matter radius of $^4$He is measured to be 1.70 ± 0.14 fm \cite{Wang2024}.  From these analyses, the $rms$ charge
radius of $^{4}$He is smaller than the $rms$ matter radius. However, the values of the $rms$ charge and matter radii are within the statistical errors. Whereas, this is an astounding puzzle, in our calculation of the folding potential we used the density that reproduces the $rms$ radii 1.70 ± 0.14 fm \cite{Wang2024}: 1.56 fm, 1.70 fm, and 1.84 fm. The latter allows us to study the influence of the $rms$ on the $\phi$-$\alpha$ potential.  
As follows from Ref. \cite{Wang2024} the simple Gaussian matter distribution model $\rho (r)=\left( \frac{C^{2}}{\pi }\right) ^{3/2}e^{-C^{2}r^{2}}$ gives $\left\langle r^{2}\right\rangle ^{1/2}=\sqrt{3/2C^{2}}$ and describes the experimental data with parameters from \cite{Wang2024}. The matter density distribution in $^4$He that leads to the experimental $rms$ within the experimental statistical uncertainty is shown in Fig. \ref{fig:05}.

\begin{table}[!ht]
\begin{center}
\caption{The parameters for the $\phi$-$N$ potential in the $^{4}S_{3/2}$ channel with statistical errors
quoted in the parentheses. In $a_{3}m_{\pi }^{4n}$ $n=1$ and $n=0$ for the $V_{\phi N}$ and $V_{G\phi N}$, respectively \cite{Lyu22}. 
The parameters for the $\alpha$-$\alpha$ potential \cite{AliBodmer}, and $\phi$-$\phi$ interaction \cite{Sofi}.}  

\label{tpp}
\begin{tabular}{ccccccc}
\hline  \noalign{\smallskip}
& \multicolumn{6}{c}{$\phi$-$N$ potential in the $^{4}S_{3/2}$ channel \cite{Lyu22}} \\ \hline  \noalign{\smallskip}
& $a_{1},$ MeV & $a_{2},$ MeV & $a_{3}m_{\pi }^{4n},$ MeV fm$^{2n}$ & $b_{1},
$ fm & $b_{2},$ fm & $b_{3},$ fm \\ \cline{1-7}  \noalign{\smallskip}
$V_{\phi N}$ & -371(27) & -119(39)  & -97(14)  & 0.13(1) & 0.30(5) & 0.63(4)
\\
$V_{G\phi N}$ & -371(19) & -50(35) & -31(53)  & 0.15(3) & 0.66(61) & 1.09(41)
\\ \hline
& \multicolumn{6}{c}{$\alpha$-$ \alpha $ potential \cite{AliBodmer} } \\ \cline{2-6} \noalign{\smallskip}
& $l$ & $V_{r},$ MeV & $V_{a},$ MeV & $\mu _{1},$ fm  & $\mu _{2},$ fm &  \\
\cline{2-6} \noalign{\smallskip}
& 0 & -30.18 & 125.0 & 2.85 & 1.53 &  \\
& 2 & -30.18 & 20.0 & 2.85 & 1.53 &  \\
& 4 & -130.0 &  & 2.11 &  &  \\ \cline{2-6}
& \multicolumn{6}{c}{$\phi$-$ \phi $ potential \cite{Sofi}} \\ \cline{3-6} \noalign{\smallskip}
&  & $V_{r},$ MeV & $V_{a},$ MeV & $\mu _{1},$ fm$^{-1}$  & $\mu _{2},$ fm$%
^{-1}$ &  \\ \cline{3-6} \noalign{\smallskip}
&  & 1000 & 1250 & 2.5 & 3 &  \\ \cline{3-6}
\end{tabular}
\end{center}
\end{table}

\section{Results of calculations and discussion}
\label{Calculations}
\subsection{Wood-Saxon type potentials for $\phi$-$\alpha$ interaction}
\label{Calculations1}
In this section we present the results of calculations for the feasibility of expected bound states for $^6_{\phi\phi}$He and $^9_\phi$Be $\phi$ mesic nuclei in the framework of a three-particle cluster model.  For  calculations of the binding energies (BE) of these systems, we use both the Wood-Saxon (WS) fit for the folding 
potential obtained based on the HAL QCD $\phi$-$ N$ potential in the $^{4}S_{3/2}$ channel \cite{Lyu22} denoted as $V_{\phi\alpha}$, and the WS potential  simulated  from the attractive potential for the $\phi$ meson in the nuclear medium originated
from the in-medium enhanced $K\bar{K}$ loop in the $\phi$-meson self-energy for three values of the cutoff
parameter $\Lambda_{K}$: 2000, 3000, and 4000 MeV \cite{Co17}, denoted as $\widetilde{V}_{\phi\alpha}$.

The input parameters for potentials are listed in Table \ref{tpp}.
We utilized the single folding model with the Gaussian mass distribution model and the HAL QCD $V_{\phi N}$ potential to develop a $V_{\phi\alpha}(r)$ potential.
Dover and Gal \cite{Gal83} proposed for the $\Xi$-$\alpha$ interaction the WS type potential. Following \cite{Gal83}, we fit the folding potential using a simple WS type expression
\begin{equation}
V_{\phi \alpha}(r)=-V_{0}\left[ 1+\exp \left( \frac{r-R}{c}\right) \right]^{-1}.
\label{Omegaalfa}
\end{equation}%
In Eq. (\ref{Omegaalfa}) $V_{0}$ is the strength of the interaction, $c$ is the surface diffuseness, and $%
R=1.1A^{1/3}$, where $A=4$ is the mass number of the nuclear core so that $R=1.74 $ fm.

\begin{figure}[ht!]
\begin{center}
\includegraphics[width=19pc]{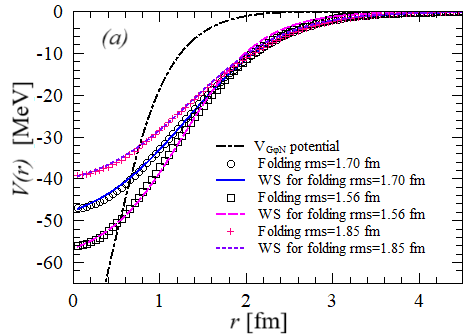}
\includegraphics[width=19pc]{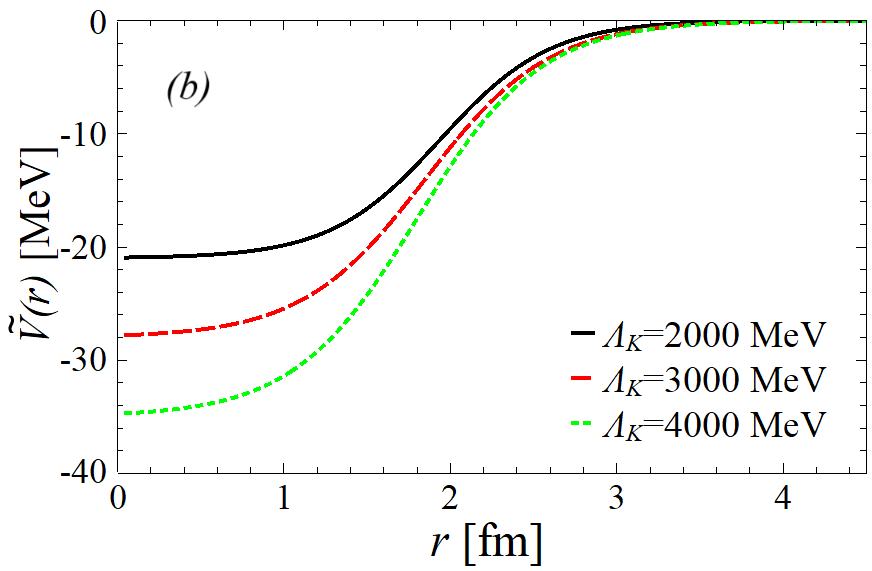}
\end{center}
\caption{($a$) The folding $\phi$-$\alpha$ potentials with the corresponding WS fits. The different symbols correspond to the different $rms$ radii. Calculations were performed for the three values of the $rms$: 1.56 fm, 1.70 fm, and 1.84 fm. The dash-dotted curve depicts the HAL QCD $\phi$-$N$ potential  \cite{Lyu22}. 
($b$) The WS type of $\phi$-$\alpha$ potential simulated from the potential proposed in Ref. \cite{Co17} for three values of the cutoff parameter $\Lambda$: 2000, 300 and 4000 MeV. }
\label{fig04}
\end{figure}

First, we present the results of calculations for the folding $\phi$-$\alpha$ potential obtained using the Gaussian density distributions given in Fig. \ref{fig:05}. The WS fit of the folding potential is denoted as $V_{\phi\alpha}$.
The dependence of the folding potential and its WS fit for different $rms$ radii is shown in Fig. \ref{fig04}$a$. One can see that
the depth of the $V_{\phi\alpha}$ potential is very sensitive to the value of
the $rms$ radius, $\it{i.e.}$ the mass density distribution, varying from $\sim -38$ MeV to $-54$ MeV. Let us mention that at the asymptotical region $r>2$ fm the density distribution drops drastically for large distances. The latter leads to the best-fitting parameter  $R=0.856A^{1/3}$ fm for $^4$He.

In Ref. \cite{Thomas97} was chosen the other density model to reproduce the $rms$ matter radius of
$^4$He, 1.56 fm, and the measured central depression in the density. Employing this density matter model the $\phi$-meson–nucleus potentials were calculated based on an effective
Lagrangian approach \cite{CoPL} using a local density approximation, with the inclusion of the $K\bar{K}$ meson loop in the
$\phi$-meson self-energy \cite{Co17}. We simulated this interaction by the WS-type potential. In Fig. \ref {fig04}$b$ depict the fit by (\ref{Omegaalfa}) of the corresponding $\phi$-$^4$He interaction for three values of the cutoff
parameter $\Lambda_{c} = 2000$, 3000, and 4000 MeV. This potential is denoted as $\widetilde{V}_{\phi \alpha }$. One can see that
the depth of the $\widetilde{V}_{\phi \alpha }$ potential is sensitive to
the cutoff parameter, varying from $\sim -20$ MeV to $-35$ MeV. This is consistence with theoretical and experimental findings that suggest of attractive $\phi$-nuclei potentials with a depth of $\sim-(20-30)$ MeV \cite{Paryev2017}.
The simulation of this $\phi$-$\alpha$ interaction by the WS type potential gives $R=1.24A^{1/3}$ fm, in contrast to $R=0.856A^{1/3}$ obtained in the case when we used the HAL QCD $V_{\phi N}$ potential.

\subsection{ $^9_\phi$Be in a three-body $\phi$+$\alpha$+$\alpha$ cluster model}
\label{sec:5a}

First, let us focus on the $^5_\phi$He nucleus within the two-body cluster model as the $\phi$+$\alpha$ system. Results of calculations for the binding energy and scattering length for the $\phi$+$\alpha$ system with $\widetilde{V}_{\phi \alpha }$ and $V_{\phi \alpha }$ potentials along with
the WS potentials parameters are presented in Table \ref{RKt00}. These potentials are distinguished by the depth of the potential, surface diffuseness, and parameter $R$. 
The $\widetilde{V}_{\phi \alpha }$ potential has $R$ of about 2 fm in contrast to $R \sim 1.4$ fm for $V_{\phi \alpha }$ but about $\sim 45\%$ weaker 
the interaction strength.  Consequently, the binding energy, $B_{\phi\alpha}$, of the $\phi$+$\alpha$ system ($^{5}_{\phi}$He) obtained using $\widetilde{V}_{\phi \alpha }$ potential is $\sim 0.8 - 5$ MeV, in contrast to $\sim 3 - 6$ MeV utilizing the $V_{\phi \alpha }$ potential.

\begin{table}[!ht]
\caption{ Parameters for the WS simulation of the $\widetilde V_{\phi\alpha}$ potential for the $\phi$ meson in the nuclear medium originated
from the in-medium enhanced $K\bar{K}$ loop in the $\phi$-meson self-energy \cite{Co17} for cutoff parameters $\Lambda_c$=2000, 3000, and 4000 MeV.
Parameters of the WS simulation of the $V_{\phi\alpha}$ potential generated through a folding procedure from the HAL QCD $\phi N$ potential \cite{Lyu22} employing the Gaussian density function giving different $rms$ radii.
 $B_{\phi\alpha}$,  $B_{\phi\alpha\alpha}$ and $B_{\phi\phi\alpha}$ are two- and three-body BE energies for the $\phi$+$\alpha $, and  $\phi$+$\alpha $+$\alpha $,  $\phi$+$\phi$+$\alpha $ systems, respectively, in MeV. $B_{\phi \phi \alpha}(V_{\phi\phi} = 0)$ is the three-body BE in MeV for the $\phi$+$\phi$+$\alpha $ system when the interaction between two $\phi$ mesons is omitted.
}
\label{RKt00}
\begin{tabular}{ccccccccc}
\hline \noalign{\smallskip}
$\phi \alpha $ potential & $rms$, fm & $V_{0}$, MeV & $R$, fm & $c$, fm & $%
B_{\phi\alpha}$ & $B_{\phi \alpha \alpha }$ & $B_{\phi \phi \alpha} $ & $B_{\phi \phi \alpha}(V_{\phi\phi} = 0)$ \\ \hline \noalign{\smallskip}
\ \ \ \ \ \ \ \ \ $\Lambda_c =2000$ MeV & 1.56 & 21 & 1.94 & 0.33 & 0.80 & 3.20 & 1.23 & 1.67\\
 $\widetilde{V}_{\phi \alpha }$ \ \ $\Lambda_c =3000$ MeV & 1.56 & 28 & 1.94 & 0.33 & 3.19 & 6.13 & 5.41 & 6.52 \\
\ \ \ \ \ \ \ \ $\Lambda_c =4000$ MeV & 1.56 & 35 & 1.80 & 0.37 & 4.71 & 9.69& 8.19 & 9.59 \\  \hline
\noalign{\smallskip}
 & 1.85  & 43 & 1.36 & 0.55 & 2.967 & 7.03& 5.01 & 6.06 \\
 $V_{\phi \alpha }$ & 1.70 & 52 & 1.30 & 0.55 & 4.780 & 9.79&  8.32 & 9.72 \\
& 1.56  & 60 & 1.26 & 0.45 & 5.976 &10.86&  10.5 & 12.2 \\ \hline
\end{tabular}
\end{table}

Now we focus on calculations of the binding energy for the $\phi$+$\alpha$+$\alpha$ system corresponding to the $^{9}_\phi$Be nucleus. For a prediction of a possible $\phi$+$\alpha$+$\alpha$ bound state we employ $\alpha$-$\alpha$ potential with orbital states $l=0,2,4$ \cite{AliBodmer}, and attractive $\widetilde{V}_{\phi \alpha }$ and $V_{\phi \alpha }$ interactions, respectively. The results of the bound state energies for this system, $B_{\phi\alpha\alpha}$, are presented in Table \ref{RKt00}. Utilization of the $\widetilde{V}_{\phi \alpha }$ potential for three values of the cutoff
parameter $\Lambda_{K}= 2000$, 3000, and 4000 MeV \cite{Co17} leads to the BE of the $\phi$+$\alpha$+$\alpha$ system 3.30, 6.13, and 9.69 MeV, respectively. Employing the stronger $V_{\phi \alpha }$ potential derived by the folding procedure of the HAL QCD interaction gives results within $\sim 7 - 11$ MeV. Therefore, the $\phi$+$\alpha$+$\alpha$ system is more strongly bound when $\phi$-$\alpha$ interaction is generated by the HAL QCD in the $^{4}S_{3/2}$ channel potential.

\subsection{ $^6_{\phi\phi}$He in the three-body $\phi$+$\phi$+$\alpha$ cluster model} \label{sec:5b}

Let us consider the calculations for the $^6_{\phi\phi}$He in the three-body cluster model. Using the $\widetilde{V}_{\phi \alpha }$ and $V_{\phi \alpha }$ potentials obtained differently, we next calculate the $\phi$+$\phi$+$\alpha$ system bound state energies. In Table \ref{RKt00} we present the BE, $B_{\phi\phi\alpha}$, by employing  $\widetilde{V}_{\phi \alpha }$ and $V_{\phi \alpha }$ potentials.
The binding energy, $B_{\phi\phi\alpha}$, is very sensitive to the value of the cutoff parameter $\Lambda_c$ and increases almost 7 times for $\Lambda_c=2000$ and 4000 MeV.
Utilizing the $V_{\phi \alpha }$  potential 
with the Gaussian density distribution leading to $rms = 1.84$ fm gives $B_{\phi\phi\alpha} = 5.01$ MeV and increases twice for the $rms=1.56$ fm, $B_{\phi\phi\alpha} = 10.5$ MeV. Interestingly enough, the BE of the  $\phi$+$\phi$+$\alpha$ system increases when the interaction between two $\phi$ mesons is omitted and the system is bound due to only $\phi\alpha$ interactions. The latter can be explained relaying to the properties of  $\phi$-$\phi$ potential. The $\phi$-$\phi$ potential constructed in \cite{Bel2008} combines a very short-range attractive core ($V_{2}\frac{e^{-\mu_{2} r} } {r}$,  $ V_{2} = 1250$ MeV, $\mu_2 = 3.0$ fm)  at distances $r < 0.4$ fm with the long-range repulsive  part ($V_{1}\frac{e^{-\mu_{1} r} }{r}$,   $ V_{1} = 1000$ MeV, $\mu_1 = 2.5$ fm). This potential effectively is repulsive. The short-range attraction does not result in a bound state, and the repulsive barrier at distances around 1 fm limits the close approach of the particle pair in the three-body system. Consequently, the attractive part of the potential is cut off, which manifests in the three-body system as an increase in binding energy when the $\phi$-$\phi$ potential is omitted.  

Note that the $\phi$+$\phi$+$\alpha$ system is more strongly bound with the $V_{\phi \alpha }$ potential generated by the folding HAL QCD $\phi$-$ N$ interaction in the $^{4}S_{3/2}$ channel.

\subsection{Discussion}
\label{sec:5d}
The obtained $V_{\phi \alpha}$ folding potential has a Woods-Saxon parameter $R \sim 1.4$ fm, while the $\widetilde{V}_{\phi \alpha}$ potential, motivated by Ref. \cite{Co17}, has $R \sim 2$ fm. We performed a test to evaluate the differences between these types of potentials. We proposed a folding procedure in which the interaction between a $\phi$ meson and a nucleon in $^4$He involves a strong repulsive core at short distances. As a model $\phi$-$N$ potential, we used the  nucleon-nucleon MT-III potential \cite{Malfliet1969} and the $\alpha$-particle density distribution, $\rho$, with the $rms$ radius of $1.70 \pm 0.14$ fm. The MT-III potential has a strong repulsive core at short distances.
We found that, in contrast to the $V_{\phi \alpha}$ potential, the model folding potentials exhibit more complicated behavior near
 the origin, including a weak repulsive component. The short-range behavior of this folding potential is similar to that of the $\phi$-$\alpha$ potential described in Ref. \cite{Co17}. Additionally, the obtained potentials have a larger Woods-Saxon parameter $R$.
Based on these findings, we make the following assumptions:

(i) Assuming a purely attractive $\phi$-$N$ potential, the folding procedure results in a potential with a Woods-Saxon parameter $R$ smaller than the $\alpha$-particle $rms$ radius of 1.7 fm. This suggests that the $\phi$ meson could be located inside the $\alpha$-particle within the $\phi$-$\alpha$ system.

(ii) Alternatively, the $\phi$ meson could be mainly distributed outside the $\alpha$-particle quantum well. In other words, the $\phi$ meson may not be confined solely to the $\alpha$-particle when using the pair potential from Ref. \cite{Co17}. In this case, the Woods-Saxon parameter $R$ would be larger than 1.7 fm, and the corresponding $\phi$-$N$ potential would likely have a repulsive core.

We conclude that attractive $\phi$-$N$ potentials produce $\phi$-$\alpha$  folding potentials with smaller Woods-Saxon parameters $R < 1.7$ fm. In contrast, $\phi$-$N$ potentials with a strong repulsive core can lead to $\phi$-$\alpha$ folding potentials characterized by a Woods-Saxon parameter $R$ equal to or larger than 1.7 fm. This assumption requires further investigation to understand realistic $\phi$-$N$ potentials with a repulsive core, as such potentials have not yet been provided.

While our results are obtained within the framework of the Faddeev equations, the most accurate approach in few-body physics, they are based on a three-body cluster model. Further investigation, including many-body calculations using realistic nucleon-nucleon and meson-nucleon interactions as an example within the no-core shell model \cite{Wirth}, is needed to confirm the feasibility of observing bound states of $^{9}_{\phi}$Be and $^{6}_{\phi\phi}$He $\phi$ mesic nuclei. At the same time, it is noteworthy that the $^8$Be core in our model exhibits strong $\alpha$-particle clustering, appearing as a resonance state of two $\alpha$ particles. Previous cluster calculations for $^9$Be ($\alpha +\alpha +n$) \cite{FSV11}, $^{12}$C (2$\alpha +\alpha$) \cite{Igor2000}, and $^9_\Lambda$Be ($\alpha +\alpha +\Lambda$) \cite{IgorF2004} nuclei have shown reliable correspondence with existing experimental data. This underscores the complex structure of nuclei with evident $\alpha$-particle clustering \cite{Okada2024}.

The best way to justify the validity of the cluster assumption is to compare the results of the proposed theoretical model with experimental data. Unfortunately, today there are no experimental data nor no-core (\textit{ab initio}) calculations conducted for the $^{9}_{\phi}$Be and $^{6}_{\phi\phi}$He $\phi$ mesic nuclei to compare with presented results. For example, for $^{6}$He nucleus to explore the structure and dynamics of the strongly correlated many-body problem \textit{ab initio} calculation using the variational Monte Carlo method \cite{Wiringa2002}, the no-core shell model \cite{Navratil2022}, and quantum Monte Carlo method  \cite{QMC2023} were utilized. However, we lack such calculations for $^{6}_{\phi\phi}$He nucleus and the use of the cluster model is a good starting point. 
Of course, description of $^{9}_{\phi}$Be and $^{6}_{\phi\phi}$He $\phi$ mesic nuclei in cluster model is a limited consideration. However, as was demonstrated in a comprehensive study of many-body correlations and $\alpha$--clustering in the ground-state and low-lying
energy continuum of the Borromean $^{6}$He nucleus \cite{Navratil2018} that it is possible
to reproduce the correct asymptotic behavior of
the $^{6}$He  wave function  in the more limited
approach  \cite{Navratil2013,FilikhinYF2014,Navratil2014}, based solely on the three-cluster $\alpha$+$n$+$n$ model. Although additional short-range six-body correlations are necessary to correctly describe also the interior
of the wave function for both the ground and scattering states \cite{Navratil2018}. 

One has to note that the folding procedure within the condition described above was successfully applied in Refs. \cite{FSVSpectroscop,FSVJPG2005}  for analysis of the $\Lambda NN \alpha$ system ($^{7}_\Lambda$He nucleus). The $\Lambda$-$\alpha$ folding potential was used in the Faddeev calculation for cluster $^{5}_\Lambda$He$+ N + N$ system. The cluster folding model also has been used in Ref. \cite{Igor2000} where the first 0$^{+}$ excited state of $^{12}$C was considered based on the Faddeev equation with averaging over $\alpha +\alpha$ resonance wave function. The results of these studies have demonstrated an agreement with experimental data and calculations in the framework of different approaches.

\section{Conclusions}
\label{sec:6}

Within the framework of the Faddeev formalism in configuration space, we investigate bound states of $^{9}_{\phi}$Be and $^{6}_{\phi\phi}$He $\phi$ mesic nuclei using a three-body cluster model for the mirror $\phi$+$\alpha$+$\alpha$ and $\phi$+$\phi$+$\alpha$ systems. To investigate the above conjecture
by utilizing our non-relativistic potential approach, we construct and employ two WS type $\phi$-$\alpha$ interactions: $V_{\phi\alpha}$ and $\widetilde{V}_{\phi \alpha}$. 
We predict the binding energy for the $^{9}_{\phi}$Be and $^{6}_{\phi\phi}$He $\phi$ mesic nuclei in the range of 1-11 MeV and 3-10 MeV, respectively.
The values of the BEs rely on the choice of $\phi$-$\alpha$ interaction.

We constructed the folding potential $V_{\phi\alpha}$ for the $\phi$-$\alpha$ interaction 
based on the underlying $\phi$-nucleon interaction. Starting from the recently proposed lattice HAL QCD $\phi N$ potential in the $^{4}S_{3/2}$ channel \cite{Lyu22} and a Gaussian form of the $\alpha$-particle matter distribution, we obtain the WS fit for the $V_{\phi\alpha}$ potential defined by Eqs. (\ref{2Dintegral})-(\ref{Omegaalfa}). Conversely, we obtained the WS simulation for the $\widetilde{V}_{\phi \alpha}$ potential based on the attractive potential for the $\phi$ meson in the nuclear medium, originating from the in-medium enhanced $K\bar{K}$ loop in the $\phi$-meson self-energy \cite{Co17}.

The $V_{\phi\alpha}$ potential leads to strongly bound $\phi$+$\alpha$+$\alpha$ and $\phi$+$\phi$+$\alpha$ systems with BEs of approximately 7-11 MeV and 5-10 MeV, respectively. The variation in energy depends on the Gaussian matter density of $^4$He that reproduces the matter $rms$ radius of this nucleus within experimental error bars. In calculations, we used the Gaussian matter densities that reproduced the $rms$ radii within the experimental uncertainty.

The $\widetilde{V}_{\phi \alpha}$ potential yields weaker bound $\phi$+$\alpha$+$\alpha$ and $\phi$+$\phi$+$\alpha$ systems with energies of approximately 3-9 MeV and 1-8 MeV, respectively. The variation in energy depends on the cutoff parameter $\Lambda_K$ \cite{Co17}. However, a comparison between both calculations reveals qualitative agreement between these approaches for a bound state in the $\phi$+$\alpha$+$\alpha$ and $\phi$+$\phi$+$\alpha$ systems.

The $V_{\phi\alpha}$ potential has a WS parameter $R \sim 1.3$ fm, which is less than the $rms$ value of 1.7 fm for $^4$He. Meanwhile, the WS parameter for the $\widetilde{V}_{\phi \alpha}$ potential is $R \sim 2$ fm, which is larger than the $rms$  value of 1.7 fm for $^4$He. This suggests that in one case, the $\phi$ meson is placed within the $\alpha$ particle, while in the case of the $\widetilde{V}_{\phi \alpha}$ potential, the $\phi$ meson is distributed throughout the quantum well, not solely within the $\alpha$ particle. Both scenarios are plausible because there is no active effect of the Pauli principle of the $\phi$ meson on nucleons in the $^8$Be or $^4$He core.

\section*{Acknowledgments}

This work is supported by the City University of New York PSC CUNY Research Award \# 66109-00 54, US National Science Foundation HRD-1345219 award, the DHS (summer research team), and the Department of Energy/National Nuclear Security Administration Award Number DE-NA0004112.

\appendix

\section{Folding potential}

Substitution of the HAL QCD $V_{\phi N}$ interaction parameterised with
three Gaussian functions (\ref{phiN}) and density \cite{Wang2024}
\begin{equation}
\rho (r)=\rho (r)=\left( \frac{C^{2}}{\pi }\right) ^{3/2}e^{-C^{2}r^{2}}
\label{RhoGauss}
\end{equation}
in (\ref{2Dintegral}) gives three integrals
\begin{equation}
F(r,u)=a_{i}\int_{0}^{\infty
}dxe^{-C^{2}x^{2}}e^{(x^{2}+r^{2}-2xru)/b_{i}^{2}}x^{2},\text{ \ \ \ \ }%
i=1,2,3.  \label{FirsInt}
\end{equation}%
The integration gives
\begin{equation}
F(r,u)=\frac{a_{i}b_{i}}{4}\frac{e^{-\frac{r^{2}}{b_{i}^{2}}}}{d_{i}^{5}}%
\times \left[ 2b_{i}d_{i}ru+\sqrt{\pi }e^{\frac{r^{2}u^{2}}{%
b_{i}^{2}d_{i}^{2}}}\left( b_{i}^{2}d_{i}^{2}+2r^{2}u^{2}\right) \left(
\text{erf}\left( \frac{ru}{\text{$b_{i}$}d_{i}}\right) +1\right) \right] ,
\label{eqn1}
\end{equation}%
where $d_{i}=\sqrt{1+C^{2}\text{$b_{i}$}^{2}}$.
Using $F(r,u)$ as the integrant in (\ref{2Dintegral}) yields
\begin{eqnarray}
V_{\phi \alpha }^{F}(r) &=&4C^{3}\exp \left( -\frac{r^{2}}{%
b_{1}^{2}+b_{2}^{2}+b_{3}^{2}}\right) \left\{ \frac{a_{1}b_{1}^{3}}{d_{1}^{3}%
}\exp \left[ \left( \frac{1}{b_{2}^{2}}+\frac{1}{b_{3}^{2}}+\frac{1}{%
b_{1}^{2}d_{1}}\right) r^{2}\right] \right.  + \notag \\
&&\frac{a_{2}b_{2}^{3}}{d_{2}^{3}}\exp \left[ \left( \frac{1}{b_{1}^{2}}+%
\frac{1}{b_{3}^{2}}+\frac{1}{b_{2}^{2}d_{2}^{2}}\right) r^{2}\right] +
\notag \\
&&\left. \frac{a_{3}b_{3}^{3}}{d_{3}^{3}}\exp \left[ \left( \frac{1}{%
b_{1}^{2}}+\frac{1}{b_{2}^{2}}+\frac{1}{b_{3}^{2}d_{3}^{2}}\right) r^{2}%
\right] \right\} .
\end{eqnarray}%

\end{document}